\documentclass[useAMS,usenatbib,onecolumn,usegraphicx]{mn2e}

\newcommand{\gsim}{\, \raisebox{-0.8ex}{$\stackrel{\textstyle >}{\sim}$ }}

\newcommand{\beq}{\begin{equation}}
\newcommand{\eeq}{\end{equation}}
\newcommand{\beqar}{\begin{eqnarray}}
\newcommand{\eeqar}{\end{eqnarray}}

\title[Possible Constraints on $L$ from QPOs in SGRs]
{Possible Constraints on the Density Dependence of the Nuclear
Symmetry Energy from Quasiperiodic Oscillations in Soft Gamma Repeaters}
\author[H. Sotani, K. Nakazato, K. Iida, \& K. Oyamatsu]
{Hajime Sotani$^1$ \thanks{E-mail:sotani@yukawa.kyoto-u.ac.jp},
Ken'ichiro Nakazato$^2$,
Kei Iida$^3$, and
Kazuhiro Oyamatsu$^4$
\\
$^1$Yukawa Institute for Theoretical Physics, Kyoto University, Kyoto 606-8502, Japan\\
$^2$Faculty of Science \& Technology, Tokyo University of Science, 2641 Yamazaki, Noda, Chiba 278-8510, Japan\\
$^3$Department of Natural Science, Kochi University, 2-5-1 Akebono-cho, Kochi 780-8520, Japan\\
$^4$Department of Human Informatics, Aichi Shukutoku University, 9 Katahira, Nagakute, Aichi 480-1197, Japan}

\begin{document}
\maketitle
\label{firstpage}

%%%%%%%%%%%%%%%%%%%%%%%%%%%%%%%%%%%%%%%%%%%%%%%%
% Abstract
\begin{abstract}
We systematically examine the fundamental frequencies of shear torsional 
oscillations in neutron star crusts in a manner that is dependent on 
the parameter $L$ characterizing the poorly known density dependence of 
the symmetry energy.  The identification of the 
lowest quasiperiodic oscillation (QPO) among the observed QPOs
from giant flares in soft-gamma repeaters as the $\ell=2$ fundamental 
torsional oscillations enables us to constrain the parameter $L$ as 
$L\ge 47.4$ MeV, which is the most conservative restriction on $L$ 
obtained in the present work that assumes that the mass and radius of 
the flaring neutron stars range 1.4-1.8 $M_{\odot}$ and 10-14 km.  
Next, we identify one by one a set of the
low-lying frequencies observed in giant flares as the fundamental torsional 
oscillations.  The values of $L$ that can reproduce
all the observed frequencies in terms of the torsional oscillations
coupled with a part of dripped neutrons via entrainment effects are then
constrained as 101.1 MeV $\le L \le$ 131.0 MeV.   Alternatively, if
only the second lowest frequency observed in SGR 1806$-$20 has a different 
origin, one obtains relatively low $L$ values ranging 58.0 MeV $\le L \le$ 
85.3 MeV, which seem more consistent with other empirical constraints 
despite large uncertainties.
\end{abstract}
%%%%%%%%%%%%%%%%%%%%%%%%%%%%%%%%%%%%%%%%%%%%%%%%


%%%%%%%%%%%%%%%%%%%%%%%%%%%%%%%%%%%%%%%%%%%%%%%%
\section{Introduction}
\label{sec:I}
%%%%%%%%%%%%%%%%%%%%%%%%%%%%%%%%%%%%%%%%%%%%%%%%

Neutron stars are believed to form as stellar remnants of core-collapse 
supernovae, which occur at the end of massive star evolution.  The 
density of matter inside a neutron star can easily become
higher than normal nuclear density, a situation hard to realize 
in the laboratory.  Due to such high densities and resulting strong 
coupling nature, the equation of state (EOS) for neutron star matter 
remains to be fixed.  Consequently, neutron stars act as a unique
laboratory to constrain the properties of such dense matter.  In 
fact, the discovery of a neutron star with $\sim 2M_\odot$ can rule out some of
soft EOSs \citep{2Msun}.  It is also suggested that information about 
neutron star interiors can be obtained via neutron star asteroseismology
or via possible detection of the spectra of gravitational waves emitted 
from neutron stars (e.g., \cite{AK1996,Sotani2001,Sotani2004,SYMT2011}). 
Unlike gravitational waves, there are some observational evidences for 
neutron star oscillations, namely, quasiperiodic oscillations 
(QPOs) in giant flares from soft gamma repeaters (SGRs).  Since SGRs are 
considered to be magnetars, which are neutron stars with strong 
surface magnetic fields of order $B \gsim 10^{14}$-$10^{15}$ G, the 
observed QPOs could be strongly associated with neutron star oscillations. 
%irrespective of what portion of a star is actually oscillating.   
So far, 
three giant flares have been detected from SGR 0526$-$66, SGR 1900+14, and 
SGR 1806$-$20, respectively.  The timing analysis of the X-ray 
afterglow of the giant flares shows that in two of them, specific oscillation 
frequencies arose in the range from tens of hertz up to a few kilohertz 
\citep{WS2006}.

After this discovery, many theoretical attempts to explain the 
observations have been made in terms of shear torsional oscillations 
in the crust of a neutron star and/or magnetic oscillations (e.g., 
\cite{Levin2006,Levin2007,Lee2007,SA2007,Sotani2007,Sotani2008a,Sotani2008b,
Sotani2009}). The important conclusion obtained from such attempts is that 
either the torsional oscillations or magnetic oscillations dominate the 
excited oscillations near the surface of a magnetized neutron star, 
depending on the magnetic field strength \citep{CK2011,CK2012,GCFMS2011,
GCFMS2012a} and magnetic configuration \citep{GCFMS2012b}. 
Given the magnetic field strength inferred from the spindown 
observations \citep{K1998,H1999}, we can reasonably consider the QPOs 
observed in giant flares as crustal torsional oscillations, 
especially in the case of low-lying modes \citep{PL2013}.  This
identification would allow one to probe the properties of matter 
in the crust, which are in turn related to the poorly known
density dependence of the symmetry energy of nuclear matter
\citep{SW2009,Sotani2011,GNHL2011,SNIO2012}.  
Once the density dependence of the symmetry energy is fixed,
furthermore, one can address the thickness of the region where nuclei 
of nonuniform pastalike structures \citep{LRP1993,O1993}
occur in a neutron star \citep{OI2007}.

      The deeper inside a neutron star, the more uncertainties in the EOS 
of neutron star matter.  Between an ionic ocean near the star's surface and a 
fluid core, a crust occurs.  In most of the crust, nuclei are 
believed to form a bcc Coulomb lattice in a roughly uniform electron sea. 
When the density becomes higher than about $4\times 10^{11}$ g cm$^{-3}$, 
neutrons start to drip out of the nuclei, and some of them are expected 
to behave as a superfluid  as long as the temperature is below the 
critical temperature.  In fact, such superfluidity plays an important 
role in modeling pulsar glitches \citep{Sauls}, while most of observed 
neutron stars are considered to be so cool that most of the dripped
neutrons would behave as a superfluid in the absence of the lattice.
According to the recent band calculations by \cite{Chamel2012}, however, 
a significant fraction of the dripped neutrons can be entrained
non-dissipatively by protons in the nuclei via Bragg scattering off the 
lattice even at zero temperature.  Meanwhile, neutron superfluidity 
due to the remaining dripped neutrons affects the crustal torsional 
oscillations, because such oscillations would be controlled by the enthalpy 
density of the constituents that comove with the protons 
\citep{vanhorn90}.  There are earlier publications that examine 
the influence of neutron superfluidity on the torsional oscillations in the 
crust for a specific crust EOS in the Newtonian framework
\citep{AGS2009,SA2009,PA2012}, while we have recently performed
systematic calculations in the relativistic framework, which were 
presented only briefly \citep{SNIO2013}. 
In this paper, we thus present
full details of such calculations, which give possible constraints on 
the parameter $L$ characterizing the density dependence of the
symmetry energy by comparing the fundamental frequencies of the shear 
torsional oscillations with the QPO frequencies observed in SGRs.
We remark that \cite{Deibel2013} performed a similar
systematic analysis independently.

     The paper is constructed as follows.  In the next section, we 
describe the crust EOSs and equilibrium configurations that will be
used for calculations of the crustal oscillation frequencies.  In the third 
section, we give the perturbation equation that governs the shear 
torsional oscillations, as well as the boundary conditions, and then
address how to take into account the effect of neutron superfluidity. 
In the fourth section, we show the numerical results for the oscillation
frequencies and possible constraints on $L$.  Finally, the paper closes 
with a conclusion.  We adopt units of $c=G=1$, where $c$ and $G$ denote the 
speed of light and the gravitational constant, respectively.

%%%%%%%%%%%%%%%%%%%%%%%%%%%%%%%%%%%%%%%%%%%%%%%%
\section{Crust Equilibrium Configuration}
\label{sec:II}
%%%%%%%%%%%%%%%%%%%%%%%%%%%%%%%%%%%%%%%%%%%%%%%%

We begin with the properties of matter in neutron star crusts, 
which are to some degree constrained by empirical data 
for nuclear masses and radii.  Near the saturation point of symmetric 
nuclear matter at zero temperature, the bulk energy per nucleon can be 
written as a function of baryon number density $n_b$ and neutron excess 
$\alpha$ \citep{L1981}:
\begin{equation}
w = w_0  + \frac{K_0}{18n_0^2}(n_b-n_0)^2 + \left[S_0 
         + \frac{L}{3n_0}(n_b-n_0)\right]\alpha^2, \label{eq:w}
\end{equation}
where $w_0$, $n_0$, and $K_0$ denote the saturation energy, saturation density,
and incompressibility of symmetric nuclear matter.   Meanwhile, $S_0$ and $L$ 
are the parameters associated with the symmetry energy coefficients $S(n_b)$, 
i.e., $S_0\equiv S(n_0)$ and $L\equiv 3n_0(dS/dn_b)$ at $n_b=n_0$. 
Among these five parameters, $w_0$, $n_0$, and $S_0$ are fairly
well constrained by empirical masses and radii of stable nuclei, while
the remaining two parameters $L$ and $K_0$ are left uncertain
\citep{OI2003}.  To see how strongly the parameters are constrained,
two of us (K.O and K.I) first constructed the model for the bulk
energy $w(n_n, n_p)$ of nuclear matter as a function of neutron and 
proton number densities, $n_n=n_b(1+\alpha)/2$ and $n_p=n_b(1-\alpha)/2$,
in such a way as to reproduce Eq.\ (\ref{eq:w}) in the limit of $n_b\to n_0$ 
and $\alpha\to 0$.  Then, within a simplified version of the extended 
Thomas-Fermi theory in which the energy density functional for a 
nucleus of neutron number $N$ and proton number $Z$,
\begin{equation}
 E=E_b+E_g+E_C+Nm_n+Zm_p,
\label{e}
\end{equation}
with the bulk energy 
\begin{equation}
  E_b=\int d^3 r \left(n_n({\bf r}) + n_p({\bf r})\right)w\left[n_n({\bf r}),n_p({\bf r})\right],
\label{eb}
\end{equation}
the gradient energy
\begin{equation}
  E_g=F_0 \int d^3 r |\nabla \left(n_n({\bf r}) + n_p({\bf r})\right)|^2,
\label{eg}
\end{equation}
the Coulomb energy
\begin{equation}
  E_C=\frac{e^2}{2}\int d^3 r \int  d^3 r' 
      \frac{n_p({\bf r})n_p({\bf r'})}{|{\bf r}-{\bf r'}|},
\label{ec}
\end{equation}
and the neutron (proton) rest mass $m_n$ ($m_p$),
was optimized with respect to the density distributions
$n_n({\bf r})$ and $n_p({\bf r})$, which are for simplicity
parametrized as
\begin{equation}
  n_i(r)=\left\{ \begin{array}{lll}
  n_i^{\rm in}\left[1-\left(\displaystyle{\frac{r}{R_i}}\right)^{t_i}\right]^3,
         & \mbox{$r<R_i,$} \\
             \\
         0,
         & \mbox{$r\geq R_i,$}
 \end{array} \right.
\label{ni}
\end{equation}
where $i=n$ and $p$.
Finally, for given $y\equiv -K_0S_0/(3n_0L)$ and $K_0$, the most relevant 
values of $w_0$, $n_0$,  $S_0$, and $F_0$ were obtained by fitting the 
charge number, mass excess, and charge radius, which can be calculated 
from the optimal density distribution, to the empirical values for
stable nuclei.   We remark that the parameter $y$ corresponds to the 
slope of the saturation line in the vicinity of $\alpha=0$ \citep{OI2003}.

     In order to obtain the equilibrium nuclear shape and size as well as 
the crust EOS for various sets of $y$ and $K_0$ at zero temperature, 
as in \cite{O1993},
\cite{OI2007} generalized Eqs.\ (\ref{e}) and (\ref{ni}) to include
dripped neutrons of uniform number density $n^{\rm out}_n$, a 
neutralizing background of electrons of uniform number density $n_e$,
and the lattice energy within a Wigner-Seitz approximation.  
As a result of optimization of the total energy density with respect 
to the parameters characterizing the nucleon distributions for given $y$ 
and $K_0$, the optimal energy density $\rho$ and nucleon 
distributions were obtained as a function of $n_b$.  
As in \cite{OI2007,SNIO2012,SNIO2013}, 
we  here confine ourselves to the parameter range $0<L<160$ MeV, 
180 MeV $ \le K_0\le$ 360 MeV, and $y<-200$ MeV fm$^3$, which equally well 
reproduce the mass and radius data for stable nuclei and effectively cover 
even extreme cases \citep{OI2003}.  The EOS parameter sets adopted in 
the present analysis are tabulated  in Table \ref{tab:EOS}, where 
the baryon number densities $n_1$ at which the nuclear shape changes 
from sphere to cylinder and $n_2$ at which the nuclear matter becomes 
uniform are also listed.  The interval between $n_1$ and $n_2$ 
corresponds to the pasta region, which decreases with $L$ and vanishes at 
$L\sim 100$ MeV \citep{OI2007}. We remark that, in order to fill gaps 
in the values of $L$, which appear in Table I in \cite{SNIO2012},
we add two more parameter sets, namely, 
$(y,K_0)=(-220~{\rm MeV~fm}^3,280~{\rm MeV})$
and $(-350~{\rm MeV~fm}^3,280~{\rm MeV})$, 
where the corresponding values of $L$ are 97.5 and 54.9 MeV.

%%%%%%%%%%%%%%%%%%%%%%%%%%%%%%%%%%%
% Table 1
%%%%%%%%%%%%%%%%%%%%%%%%%%%%%%%%%%%
\begin{table*}
\centering
 \begin{minipage}{98mm}
\caption{The EOS parameters adopted in the present analysis and 
the corresponding lower and upper densities of the pasta region. That is,
 $n_1$ denotes the baryon number density at which
the nuclear shape changes from sphere to cylinder, while $n_2$
denotes that at which the nuclear matter becomes uniform.
}
\begin{tabular}{ccccccc}
\hline\hline
 & $y$ (MeV fm$^3$) & $K_0$ (MeV) & $L$ (MeV) & $n_1$ (fm$^{-3}$) & $n_2$ (fm$^{-3}$) & \\
\hline
 &   $-220$ & 180 & 52.2 & 0.060 & 0.079 &  \\
 &   $-220$ & 230 & 73.4 & 0.064 & 0.073 &  \\
 &   $-220$ & 280 & 97.5 & 0.067 & 0.068 &  \\
 &   $-220$ & 360 & 146.1 & 0.066 & 0.066 &  \\
 &   $-350$ & 180 & 31.0 & 0.058 & 0.091 &  \\
 &   $-350$ & 230 & 42.6 & 0.063 & 0.086 &  \\
 &   $-350$ & 280 & 54.9 & 0.067 & 0.083 &  \\
 &  $-350$ & 360 & 76.4 & 0.072 & 0.076 &  \\
 & $-1800$ & 180 & 5.7 & 0.058 & 0.134 &  \\
 & $-1800$ & 230 & 7.6 & 0.058 & 0.127 &  \\
%  &-1800 & 280 & *** & *** & *** &  \\
 & $-1800$ & 360 & 12.8 & 0.058 & 0.118 &  \\
\hline\hline
\end{tabular}
\label{tab:EOS}
\end{minipage}
\end{table*}
%%%%%%%%%%%%%%%%%%%%%%%%%%%%%%%%%%%

Let us now consider the equilibrium neutron star configurations.
Since the magnetic energy is much smaller than the gravitational binding 
energy even for magnetars, we can neglect the deformation due to the 
magnetic pressure. Additionally, since the magnetars are relatively 
slowly rotating, we can also neglect the rotational effect.  Hereafter,
therefore, we consider spherically symmetric neutron stars, 
whose structure is described by the solutions of the well-known 
Tolman-Oppenheimer-Volkoff (TOV) equations.  In this case, the metric 
can be expressed in terms of the spherical polar coordinates $r$, 
$\theta$, and $\phi$ as
\begin{equation}
 ds^2 = -{\rm e}^{2\Phi}dt^2 + {\rm e}^{2\Lambda}dr^2 + r^2 d\theta^2 
+ r^2\sin^2\theta\, d\phi^2,
\label{metric}
\end{equation}
where $\Phi$ and $\Lambda$ are functions of $r$. We remark that $\Lambda(r)$ 
is associated with the mass function 
\begin{equation}
m(r)=\int_0^r dr' 4\pi r'^2\rho (r') 
\end{equation}
as ${\rm e}^{2\Lambda}=(1-2m(r)/r)^{-1}$.

     To solve the TOV equations, one generally uses the 
zero-temperature EOS, i.e., the pressure $p$ as a function of the 
energy density $\rho$.  For matter in the crust, we use the same EOS 
models as described above.  Unfortunately, the core EOS  is still
uncertain in the absence of clear understanding of the constituents and 
their interactions both in vacuum and in medium. Since we will focus on  
shear torsional oscillations that occur in the crust, we can effectively
describe such uncertainties in the core EOS by solely setting the star's mass 
$M$ and radius $R$ as free parameters, without using specific models for 
the core EOS.  In fact, for various sets of $M$ and $R$, we 
systematically construct the equilibrium configuration of the crust 
by integrating the TOV equations with the crust EOS from the  star's 
surface all the way down to the crust-core boundary as in 
\cite{IS1997,SNIO2012,SNIO2013}.  This is a contrast to the usual way 
of constructing a star by initially giving a value of the central mass 
density and then integrating the TOV equations with a specific model for
the EOS from the star's center to surface.   Hereafter, we will consider 
$1.4\le M/M_\odot\le 1.8$ and 10 km $\le R\le$ 14 km as typical values 
of $M$ and $R$.  Such choice of $M$ and $R$ can duly encapsulate 
uncertainties of the core EOS.

    Generally, a restoring force for shear torsional oscillations is 
provided by shear stress,  which comes from the elasticity of the
oscillating body and is characterized by the shear modulus $\mu$.  In 
the case of torsional oscillations in the crust of a neutron star, the 
shear modulus is determined by the lattice energy of the Coulomb crystal
that constitutes the crust.  Since the crystal is generally considered to be 
of bcc type (an fcc lattice might occur in place of bcc in the crust
as suggested by recent Thomas-Fermi calculations \citep{Okamoto2012}), one can 
use the corresponding shear modulus, which is calculated for $Ze$ point 
charges of number density $n_i$ as
\begin{equation}
\mu = 0.1194\times\frac{n_i (Ze)^2}{a}, \label{eq:shear}
\end{equation}
where $a=(3Z/4\pi n_e)^{1/3}$ is the radius of a Wigner-Seitz cell 
\citep{SHOII1991}.  Note that this formula is derived 
in the limit of zero temperature from Monte Carlo calculations for 
the shear modulus averaged over all directions \citep{OI1990}. 
As shown in \cite{SNIO2012}, the shear modulus depends strongly on the 
value of $L$, which comes mainly from the $L$ dependence of the 
calculated $Z$ \citep{OI2007}.  It is natural that one should take 
into account the shear modulus in pasta phases, if present, but
hereafter we simply assume $\mu=0$ for the pasta phases, as in 
\cite{GNHL2011,SNIO2012,SNIO2013}.  This is because the shear modulus in 
the pasta phases except a phase of spherical bubbles has at least one 
direction in which the system is invariant with respect to translation
and hence is expected to be significantly smaller than that in a phase 
of spherical nuclei \citep{PP1998}.  Under this assumption,  we have 
only to consider the shear torsional oscillations that are excited 
within a crustal region of spherical nuclei, i.e., for $n_b\le n_1$. 
Anyway, the constraint on $L$ that will be given below 
can be considered to be robust, because the pasta region is highly limited 
given the resulting constraint on $L$ \citep{OI2007}.

%%%%%%%%%%%%%%%%%%%%%%%%%%%%%%%%%%%%%%%%%%%%%%%%
\section{Torsional Oscillations}
\label{sec:III}
%%%%%%%%%%%%%%%%%%%%%%%%%%%%%%%%%%%%%%%%%%%%%%%%

     We  now consider the shear torsional oscillations on the equilibrium 
configuration of the crust of a neutron star described above. In order to
determine the frequencies, we adopt the relativistic Cowling approximation, 
i.e., we neglect the metric perturbations on Eq.\ (\ref{metric}) by 
setting $\delta g_{\mu\nu}=0$.  In fact, one can consider the shear torsional 
oscillations with satisfactory accuracy even with the relativistic Cowling 
approximation, because the shear torsional oscillations on a spherically 
symmetric star are incompressible and thus independent of the density 
variation during such oscillations.  Additionally, due to the spherically 
symmetric nature of the background, we have only to consider the 
axisymmetric oscillations.  Then, the only non-zero perturbed matter 
quantity is the $\phi$ component of the perturbed four-velocity, 
$\delta u^{\phi}$, which can be written as
\begin{equation}
   \delta u^{\phi} = {\rm e}^{-\Phi}\partial_t {\cal Y}(t,r)\frac{1}{\sin\theta}\partial_\theta P_{\ell}(\cos\theta), 
\end{equation}
where $\partial_t$ and $\partial_\theta$ denote the partial derivatives with 
respect to $t$ and $\theta$, respectively, while $P_\ell(\cos\theta)$ is 
the $\ell$-th order Legendre polynomial.  We remark that ${\cal Y}(t,r)$ 
characterizes the radial dependence of the angular displacement of a
matter element. By assuming that the perturbation variable ${\cal Y}(t,r)$ has
such a harmonic time dependence as ${\cal Y}(t,r)={\rm e}^{{\rm i}\omega 
t}{\cal Y}(r)$, the perturbation equation that governs the shear 
torsional oscillations can be derived from the linearized equation of 
motion as  \citep{ST1983}
\begin{equation}
 {\cal Y}'' + \left[\left(\frac{4}{r}+\Phi'-\Lambda'\right)+\frac{\mu'}{\mu}\right]{\cal Y}'  %\nonumber \\
      + \left[\frac{H}{\mu}\omega^2{\rm e}^{-2\Phi}-\frac{(\ell+2)(\ell-1)}{r^2}\right]{\rm e}^{2\Lambda}{\cal Y} = 0,
 \label{eq:perturbation}
\end{equation}
where $H$ is the enthalpy density defined as $H\equiv \rho+p$ with the 
energy density $\rho$ and pressure $p$ as described in Sec.\ 
\ref{sec:II}, and the prime denotes the derivative with respect to $r$.

     Once appropriate boundary conditions are imposed, the problem to 
solve reduces to an eigenvalue problem with respect to $\omega$.  Since 
there is no matter outside the star, we adopt the zero-torque condition at the 
star's surface.  Meanwhile, since there is no traction force in the 
region with $\mu=0$, we adopt the zero-traction condition at the position 
where spherical nuclei disappear in the deepest region of the crust.  In 
practice, one can show that both conditions reduce to ${\cal Y}'=0$ 
\citep{ST1983,Sotani2007}.  Thus, in determining the frequencies of 
the shear torsional oscillations, we impose the condition of 
${\cal Y}'=0$ both at $n_b=0$ and $n_1$.

     Now, we take into account the effect of neutron superfluidity on the 
shear torsional oscillations.  In general, it is considered that neutrons 
confined in the nuclei start to drip therefrom when the mass 
density becomes more than $\sim 4\times 10^{11}$ g cm$^{-3}$.  Then, some of 
the dripped neutrons can behave as a superfluid.  Although the behavior of the 
dripped neutrons is not fully understood, a significant fraction of 
the dripped neutrons may move non-dissipatively with protons in the 
nuclei as a result of Bragg scattering off the bcc lattice of the nuclei.
In fact, the recent band calculations beyond the Wigner-Seitz approximation 
by \cite{Chamel2012} show that the superfluid density, which is 
defined here as the density of neutrons unlocked to the motion of protons in 
the nuclei, depends sensitively on the baryon density above neutron drip
and that a considerable portion of the dripped neutrons can be 
locked to the motion of protons in the nuclei.  On the other hand, since the 
shear torsional oscillations are transverse, the remaining superfluid 
neutrons, whose low-lying excitations are longitudinal, do not contribute
to such oscillations \citep{PCR2010}.

     We build the effect of neutron superfluidity into the 
effective enthalpy density $\tilde{H}$, which can be determined by subtracting 
the superfluid mass density from the total enthalpy density $H$ in Eq.\ 
(\ref{eq:perturbation}) that fully contains the contributions of 
the superfluid neutrons as well as the nuclei and companions
\citep{IB-II}.  Since we assume that the temperature of neutron star 
matter is zero, the baryon chemical potential $\mu_b$ can be expressed as 
$\mu_b=H/n_b$.  Thus, one can write down \citep{SNIO2013}
\begin{equation}
   \tilde{H} = \left(1-\frac{N_s}{A}\right)H,
\end{equation}
where $N_s$ denotes the number of neutrons in a Wigner-Seitz cell that do not 
comove with protons in the nucleus, while $A$ is the total nucleon 
number in the Wigner-Seitz cell. Finally, substituting $\tilde{H}$ 
for $H$ in Eq.\ (\ref{eq:perturbation}), one can determine the frequencies of 
the shear torsional oscillations, which include the effect of 
neutron superfluidity in a manner that depends on the value of $N_s$. 
Hereafter, we will assume that $N_s$ comes entirely from the dripped neutron 
gas.  Even so, it is still uncertain how much fraction of the
dripped neutrons behave as a superfluid. Thus, as in \cite{SNIO2013}, we 
introduce a new parameter $N_s/N_d$, where $N_d$ is the number of the
dripped neutrons in the Wigner-Seitz cell.  For $N_s/N_d=0$, 
all the dripped neutrons behave as normal matter and contribute to 
the shear motion, while for $N_s/N_d=1$, all the dripped neutrons 
behave as a superfluid.  We remark that $N_d-N_s$ denotes the 
number of the dripped neutrons bound to the nucleus. 
Typically, the value of $N_s/N_d$ depends on the density inside a 
neutron star \citep{Chamel2012}, but the case of $N_s/N_d=0$ 
in the whole crust is closer to the typical behavior than the case of
$N_s/N_d=1$.

\begin{figure}
\begin{center}
\includegraphics[scale=0.5]{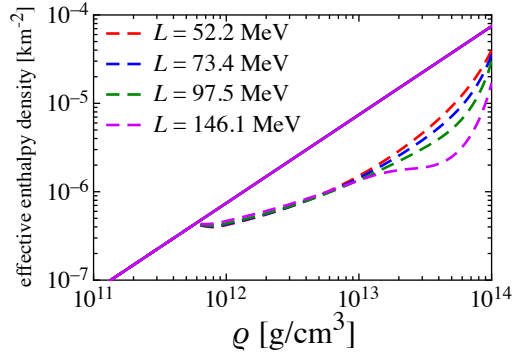} 
\end{center}
\caption{%%
(Color online) Effective enthalpy density $\tilde{H}$ calculated as a 
function of mass density $\rho$ for $y=-220$ MeV fm$^3$.  The solid
lines are the results for $N_s/N_d=0$, while the broken lines are for 
$N_s/N_d=1$.  The four broken lines from top to bottom correspond to the cases 
of $L=52.2$, 73.4, 97.5 and 146.1 MeV.
}%%
\label{fig:enthalpy}
\end{figure}

\begin{figure}
\begin{center}
\includegraphics[scale=0.5]{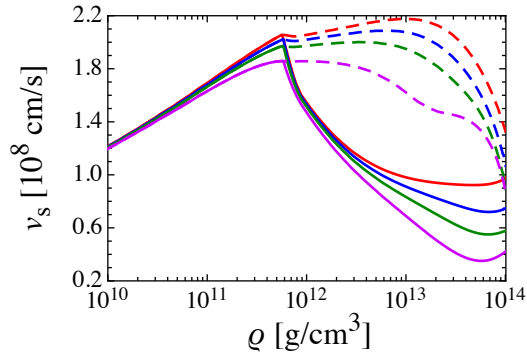} 
\end{center}
\caption{%%
(Color online) Same as Fig.\ \ref{fig:enthalpy}, but for the shear 
velocity $v_s$. 
}%%
\label{fig:vs}
\end{figure}

     In Fig.\ \ref{fig:enthalpy}, we show the effective enthalpies for 
$N_s/N_d=0$ by the solid lines and for $N_s/N_d=1$ by the broken 
lines, where we adopt the EOSs with $y=-220$ MeV fm$^3$ and with 
$L=52.2$, 73.4, 97.5 and 146.1 MeV.  From this figure, one can see that the 
effective enthalpies for $N_s/N_d=0$ are almost independent of the EOS 
parameters, while those for $N_s/N_d=1$ depend strongly on the EOS parameters 
especially for $\rho\gsim 10^{13}$ g/cm$^3$.  This is because for larger  $L$, the 
symmetry energy at subnuclear densities becomes smaller, leading to increase 
in the density of the dripped neutrons.   Another important quantity
that characterizes the shear torsional oscillations is the shear velocity 
$v_s$, which is defined as $v_s^2\equiv \mu/\tilde{H}$.  In Fig.\ \ref{fig:vs},
we depict the shear velocity with the same sets of the EOS 
parameters, where the solid lines are for $N_s/N_d=0$ and the broken lines are 
for $N_s/N_d=1$. From this figure, we find that, depending on $N_s/N_d$, the 
shear velocity can double that for $N_s/N_d=0$.  Both Figs.\ 
\ref{fig:enthalpy} and \ref{fig:vs} suggest the necessity of introducing
the effect of neutron superfluidity.

%%%%%%%%%%%%%%%%%%%%%%%%%%%%%%%%%%%
% Figure 2
%%%%%%%%%%%%%%%%%%%%%%%%%%%%%%%%%%%
%%%%%%%%%%%%%%%%%%%%%%%%%%%%%%%%%%%

%%%%%%%%%%%%%%%%%%%%%%%%%%%%%%%%%%%
% Figure 3
%%%%%%%%%%%%%%%%%%%%%%%%%%%%%%%%%%%
%%%%%%%%%%%%%%%%%%%%%%%%%%%%%%%%%%%

%%%%%%%%%%%%%%%%%%%%%%%%%%%%%%%%%%%%%%%%%%%%%%%%
\section{Constraints on the EOS parameters}
\label{sec:IV}
%%%%%%%%%%%%%%%%%%%%%%%%%%%%%%%%%%%%%%%%%%%%%%%%

\begin{figure}
\begin{center}
\includegraphics[scale=0.5]{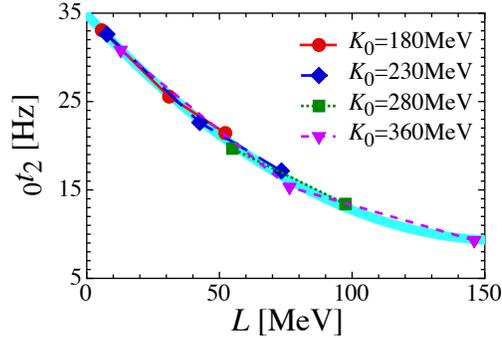} 
\end{center}
\caption{
(Color online) The $\ell=2$ fundamental frequencies of the shear 
torsional oscillations, ${}_0t_2$, which are plotted as a function of 
$L$ for $M=1.4M_\odot$ and $R=12$ km.  The thick solid line denotes the 
fitting formula (\ref{eq:fit-0t2}).
}
\label{fig:0t2-M14R12}
\end{figure}

     The shear torsional oscillations are often referred to as $t$-modes, 
which are labelled as  ${}_nt_\ell$, where $\ell$ is the angular index 
and $n$ is the number of radial nodes in the eigenfunctions of the overtones 
for a specific $\ell$.  In order to see the dependence of the shear torsional 
oscillations on the EOS parameters, we firs consider the case in 
which the effect of neutron superfluidity is ignored, i.e., $N_s/N_d=0$.
We calculate the fundamental frequencies of such oscillations for a typical 
neutron star model with $M=1.4M_\odot$ and $R=12$ km by using the 11 
EOS parameter sets shown in Table \ref{tab:EOS}.  The calculated fundamental 
frequencies with $\ell=2$ are shown in Fig.\ \ref{fig:0t2-M14R12} as a function
of $L$.  From this figure, as in \cite{SNIO2012}, one can see that the 
$\ell=2$ fundamental frequency of the shear torsional oscillations is 
almost independent of the incompressibility $K_0$, once the stellar 
model is fixed at $M=1.4M_\odot$ and $R=12$ km.  So is it for different 
stellar models with $1.4M_\odot \le M\le 1.8M_\odot$ and $10$ km $\le R\le$ 
14 km.  Thus, we can focus on the $L$ dependence of the $\ell=2$ fundamental 
frequencies of the shear torsional oscillations.

     Since the number of the EOS parameter sets is limited to 
eleven, we want to see the $L$ dependence in a continuous manner.  To 
this end, we derive a fitting formula for ${}_0t_2$ by assuming the 
polynomial function form
\begin{equation}
  {}_0t_2 = c_2^{(0)} - c_2^{(1)} L + c_2^{(2)} L^2, \label{eq:fit-0t2}
\end{equation}
where $c_2^{(0)}$, $c_2^{(1)}$, and $c_2^{(2)}$ are the adjustable positive 
parameters that depend on $M$ and $R$.  In practice, we adopt the 
Levenberg-Marquardt algorithm to derive the coefficients $c_2^{(0)}$, 
$c_2^{(1)}$, and $c_2^{(2)}$.  The obtained fitting formula is also shown in 
Fig.\ \ref{fig:0t2-M14R12} with thick solid line.  Additionally, for the 
stellar model with $M=1.4M_\odot$ and $R=12$ km, we list the 
calculated $\ell=2$ fundamental frequencies, ${}_0t_2^{(c)}$, the 
expected values from the fitting formula (\ref{eq:fit-0t2}), ${}_0t_2^{(e)}$, 
and the relative errors defined as 
$({}_0t_2^{(c)}-{}_0t_2^{(e)})/{}_0t_2^{(c)}$ in Table \ref{tab:0t2-M14R12}. 
For other stellar models, the relative errors are 
similar to the case of $M=1.4M_\odot$ and $R=12$ km.  This means that 
the fitting formula is in good agreement with the calculated frequencies at 
least within the accuracy of $\sim 5\%$.

%%%%%%%%%%%%%%%%%%%%%%%%%%%%%%%%%%%
% Figure 4
%%%%%%%%%%%%%%%%%%%%%%%%%%%%%%%%%%%
%%%%%%%%%%%%%%%%%%%%%%%%%%%%%%%%%%%

%%%%%%%%%%%%%%%%%%%%%%%%%%%%%%%%%%%
% Table 2
%%%%%%%%%%%%%%%%%%%%%%%%%%%%%%%%%%%
\begin{table*}
\centering
 \begin{minipage}{124mm}
\caption{The calculated frequencies, ${}_0t_2^{(c)}$, for the stellar model 
of $M=1.4M_\odot$ and $R=12$ km in the absence of the effect of neutron 
superfluidity, and the expected values from Eq.\ (\ref{eq:fit-0t2}), 
${}_0t_2^{(e)}$.  The relative errors, which are determined by 
$({}_0t_2^{(c)}-{}_0t_2^{(e)})/{}_0t_2^{(c)}$, are also tabulated.
}
\begin{tabular}{cccccccc}
\hline\hline
 & $y$ (MeV fm$^3$) & $K_0$ (MeV) & $L$ (MeV) & ${}_0t_2^{(c)}$ (Hz) & ${}_0t_2^{(e)}$ (Hz) & relative error (\%) & \\
\hline
 &   $-220$ & 180 & 52.2 & 21.44 & 20.73 & 3.31 &  \\
 &   $-220$ & 230 & 73.4 & 25.56 & 25.78 & $-0.86$ &  \\
 &   $-220$ & 280 & 97.5 & 33.07 & 33.04 & 0.10 &  \\
 &   $-220$ & 360 & 146.1 & 17.15 & 16.60 & 3.17 &  \\
 &   $-350$ & 180 & 31.0 & 22.63 & 22.91 & $-1.24$ &  \\
 &   $-350$ & 230 & 42.6 & 32.62 & 32.43 & 0.57 &  \\
 &   $-350$ & 280 & 54.9 & 13.41 & 13.02 & 2.96 &  \\
 &  $ -350$ & 360 & 76.4 & 19.68 & 20.16 & $-2.43$ &  \\
 & $-1800$ & 180 & 5.7 & 9.29 & 9.40 & $-1.13$ &  \\
 & $-1800$ & 230 & 7.6 & 15.32 & 16.09 & $-5.00$ &  \\
 & $-1800$ & 360 & 12.8 & 30.82 & 30.86 & $-0.14$ &  \\
\hline\hline
\end{tabular}
\label{tab:0t2-M14R12}
\end{minipage}
\end{table*}
%%%%%%%%%%%%%%%%%%%%%%%%%%%%%%%%%%%

\begin{figure}
\begin{center}
\includegraphics[scale=0.5]{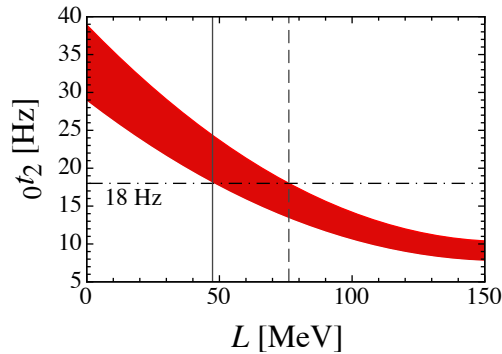} 
\end{center}
\caption{
(Color online) Expected values of ${}_0t_2$ from fitting formula 
(\ref{eq:fit-0t2}) for the stellar models with $10$ km $\le R \le 14$ km and 
$1.4M_\odot\le M\le 1.8M_\odot$.  The horizontal dot-dashed line denotes the 
lowest QPO frequency observed from SGR 1806$-$20 (Watts \& Strohmayer 2006), 
while the vertical solid and broken lines correspond $L=47.4$ MeV and $L=76.2$ 
MeV. 
}
\label{fig:fit-0t2}
\end{figure}

     In Fig.\ \ref{fig:fit-0t2}, we illustrate the values of ${}_0t_2$ 
given by the fitting formula (\ref{eq:fit-0t2}) for the stellar models 
with $10$ km $\le R \le 14$ km and $1.4M_\odot\le M\le 1.8M_\odot$, 
together with the lowest QPO frequency observed from SGR 1806$-$20 
shown in horizontal dot-dashed line.  We remark that ${}_0t_2$ decreases 
with $R$ and $M$.   In Fig.\ \ref{fig:fit-0t2}, therefore, the upper 
(lower) boundary of the painted region corresponds to the $\ell=2$ 
fundamental frequency for $M=1.4M_\odot~(1.8M_\odot)$ and $R=10~(14)$ km.
Now, on the assumption that the QPOs observed in SGR giant flares come 
from the crustal torsional oscillations, ${}_0t_2$ should become equal to
or even lower than the lowest frequency in the observed QPOs, because ${}_0t_2$
is theoretically the lowest frequency among many eigenfrequencies of the 
torsional oscillations.  Then, from Fig.\ \ref{fig:fit-0t2}, we can constrain 
$L$ as $L\ge 47.4$ MeV if the central objects of SGRs are neutron stars
with $R\le 14$ km and $M\le 1.8M_\odot$.  If the oscillating neutron star is a 
typical one with $R=10$ km and $M=1.4M_\odot$, we could make a 
severer constraint on $L$ as $L\ge 76.2$ MeV.  It should be noticed that we 
omit the effect of the pasta phases on the shear modulus in this 
analysis.  Given the resulting constraint on $L$, the pasta region is
too limited to have significant consequence to ${}_0t_2$.  In 
practice, even if we allow for the effect of the pasta phases, the 
frequency would increase because the shear modulus would effectively
increase, i.e., the painted region in Fig.\ \ref{fig:fit-0t2} would
shift to right.  Thus, our constraint on $L$ is still satisfied.  
Moreover, even if we take into account the effect of neutron superfluidity, 
the obtained constraint on $L$ would hold, because the frequencies 
would increase due to the effect of neutron superfluidity \citep{SNIO2013}.

%%%%%%%%%%%%%%%%%%%%%%%%%%%%%%%%%%%
% Figure 5
%%%%%%%%%%%%%%%%%%%%%%%%%%%%%%%%%%%
%%%%%%%%%%%%%%%%%%%%%%%%%%%%%%%%%%%

     Next, we consider the effect of neutron superfluidity on
the torsional oscillations and try to constrain $L$ by fitting 
the predicted fundamental frequencies of the shear torsional oscillations
with different values of $\ell$ to the QPO frequencies observed in SGRs.  To 
take into account the effect of neutron superfluidity, as mentioned before, 
one needs to know how much fraction of the dripped neutrons behave as 
a superfluid.  In fact, earlier calculations of such a fraction 
are very limited, while the behavior of unbound neutrons inside a nucleus
remains to be fully understood.  In this paper we adopt the result 
for $N_s/N_d$ derived by \cite{Chamel2012} as $n_n^c/n_n^f$,
which is based on the band calculations.  
According to his data, the value of $N_s/N_d$ depends on the baryon
density and becomes around $10-30$ \% at $n_b\sim 0.01-0.4n_0$. 
We remark that this value of $N_s/N_d$ does
not allow for the possible dependence on $L$, which
is still uncertain.
Using
such a value of $N_s/N_d$, we have calculated the fundamental 
frequencies of the shear torsional oscillations with different values of 
$\ell$ for the stellar models constructed with the EOS parameter sets as
shown in Table \ref{tab:EOS}.  Then, we again find that, for each 
stellar model, the calculated frequencies show negligible 
dependence on $K_0$ but are sensitive to $L$ as in the case of the 
calculations of ${}_0t_2$ ignoring the superfluid effect.

To express the calculated ${}_0t_\ell$ as a continuous function of $L$, 
we use the same form of fitting formula as Eq.\ (\ref{eq:fit-0t2}):
\begin{equation}
  {}_0t_\ell = c_\ell^{(0)} - c_\ell^{(1)} L + c_\ell^{(2)} L^2, \label{eq:fit-0tn}
\label{eq:fit-0t2e}
\end{equation}
where $c_\ell^{(0)}$, $c_\ell^{(1)}$, and $c_\ell^{(2)}$ are the adjustable 
positive parameters that depend on $M$ and $R$ for a specific index 
$\ell$.  We find that, just like the case of Eq.\ (\ref{eq:fit-0t2}),
Eq.\ (\ref{eq:fit-0t2e}) reproduces the calculated frequencies with 
sufficient accuracy to be used below for reasonable fitting to the observed
QPO frequencies.  Hereafter, we will thus refer to the expected frequencies 
from Eq.\ (\ref{eq:fit-0t2e}) as the calculated frequencies for simplicity.  
For comparison of the calculated frequencies with the observed QPO
frequencies, we particularly focus on the observed QPO frequencies 
lower than 100 Hz, i.e., 18, 26, 30, and 92.5 Hz in SGR 1806$-$20 and 28, 54, 
and 84 Hz in SGR 1900+14 \citep{WS2006}, because the higher observed 
frequencies would be easier to explain in terms not only of 
multipolar fundamental and overtone frequencies of shear torsional 
oscillations in the crust, but also of polar type 
oscillations.  Additionally, the possibility to explain the higher QPO 
frequencies in terms of shear torsional oscillations in the 
hadron-quark mixed phase that may occur in the core of a neutron 
star is also suggested by \cite{SMT2012}.

\begin{figure}
\begin{center}
\includegraphics[scale=0.5]{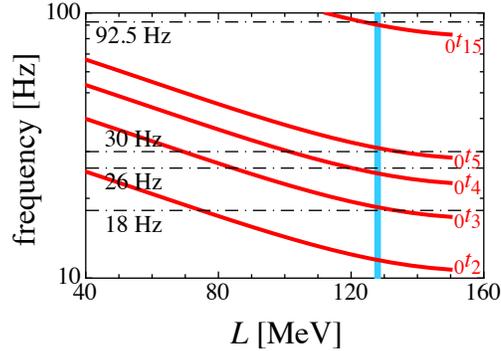} 
\end{center}
\caption{
(Color online) 
Comparison of the calculated fundamental frequencies of the shear 
torsional oscillations (solid lines) with the QPO frequencies observed in SGR 
1806$-$20 (dot-dashed lines), where we adopt the stellar model with 
$M=1.4M_\odot$ and $R=12$ km and the Chamel data for $N_s/N_d$.  The 
vertical line denotes the value of $L$ that is consistent with the 
observations.
}
\label{fig:L1-1806-M14R12}
\end{figure}

\begin{figure}
\begin{center}
\includegraphics[scale=0.5]{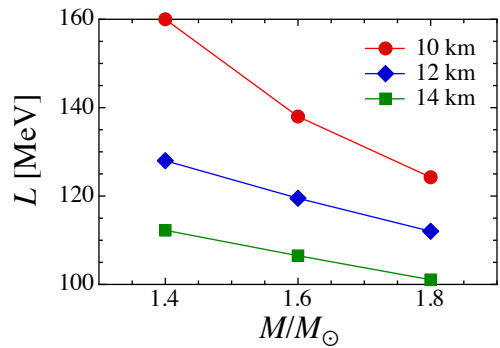} 
\end{center}
\caption{
(Color online) Values of $L$ at which the calculated fundamental frequencies 
of the shear torsional oscillations agree best with the QPO frequencies 
observed in SGR 1806$-$20.  The circles, diamonds, and squares correspond to 
the stellar models with $R=10,$ 12, and 14 km, respectively.
}
\label{fig:1806-L1}
\end{figure}

      Due to the small interval between the observed frequencies 26 and 
30 Hz in SGR 1806$-$20, theoretical explanations of the QPO frequencies 
observed in SGR 1806$-$20 are more difficult than those in SGR 
1900+14 \citep{Sotani2007}. Therefore, we first try to reproduce the 
QPOs observed in SGR 1806$-$20 by the crustal shear modes.  To find a 
full correspondence of the shear torsional oscillations to the 
observed QPOs, one should identify the lowest frequencies in SGR 1806$-$20 
(18 Hz) as the $\ell=3$ fundamental frequency as in \cite{Sotani2011,SNIO2013}.
Then, one can manage to explain 26, 30, and 92.5 Hz in terms of the fundamental
frequencies with $\ell=4$, 5, and 15.  In Fig.\ \ref{fig:L1-1806-M14R12},
we compare the predicted frequencies with the QPO frequencies observed in SGR 
1806$-$20 for a typical neutron star model with $M=1.4M_\odot$ and 
$R=12$ km.  The best value of $L$ to reproduce the observed frequencies is 
$L=128.0$ MeV, for which the calculated frequencies with the best 
value of $L$ and their relative errors from the observed 
values are shown in Table \ref{tab:1806-1}. We note that the
relative errors are of the order of such relative errors involved in using
the fitting formula (\ref{eq:fit-0t2e}) as shown in Table 
\ref{tab:0t2-M14R12}.  After performing a
similar analysis for different stellar models, we find that the QPO 
frequencies observed in SGR 1806$-$20 can be explained in terms of the 
eigenfrequencies with the same multipole fundamental oscillations even for the 
different stellar models, while the corresponding relative errors 
are similar to those shown in Table \ref{tab:1806-1}. For each 
stellar model, the obtained best value of $L$ is shown in Fig.\ 
\ref{fig:1806-L1}.  Assuming that the mass and radius of the 
oscillating neutron star are in the range $1.4\le M/M_\odot \le 1.8$ 
and 10 km $\le R \le$ 14 km, one can constrain $L$ as 
$101.1$ MeV $\le L \le 160.0$ MeV from the observed QPO frequencies of 
SGR 1806$-$20.

%%%%%%%%%%%%%%%%%%%%%%%%%%%%%%%%%%%
% Figure 6
%%%%%%%%%%%%%%%%%%%%%%%%%%%%%%%%%%%
%%%%%%%%%%%%%%%%%%%%%%%%%%%%%%%%%%%

%%%%%%%%%%%%%%%%%%%%%%%%%%%%%%%%%%%
% Table 3
%%%%%%%%%%%%%%%%%%%%%%%%%%%%%%%%%%%
\begin{table*}
\centering
 \begin{minipage}{78mm}
   \caption{The QPO frequencies observed in SGR 1806$-$20 and the
calculated frequencies with the best value of $L$ to reproduce the 
observed values for the stellar model with $M=1.4M_\odot$ and $R=12$ km.
}
\begin{tabular}{cccccc}
\hline\hline
 &  QPO frequency (Hz) & $\ell$ & ${}_0t_\ell$ (Hz) & relative
error (\%) & \\
\hline
  & 18 & 3 & 18.50 & $-2.79$ &  \\
  & 26 & 4 & 24.82 & 4.53 &  \\
  & 30 & 5 & 30.96 & $-3.19$ &  \\
  & 92.5 & 15 & 90.18 & 2.51 &  \\
\hline\hline
\end{tabular}
\label{tab:1806-1}
\end{minipage}
\end{table*}
%%%%%%%%%%%%%%%%%%%%%%%%%%%%%%%%%%%

%%%%%%%%%%%%%%%%%%%%%%%%%%%%%%%%%%%
% Figure 7
%%%%%%%%%%%%%%%%%%%%%%%%%%%%%%%%%%%
%%%%%%%%%%%%%%%%%%%%%%%%%%%%%%%%%%%

\begin{figure}
\begin{center}
\includegraphics[scale=0.5]{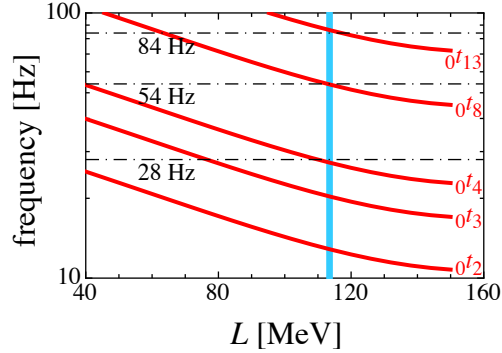} 
\end{center}
\caption{
(Color online) Same as Fig.\ \ref{fig:L1-1806-M14R12}, but for
comparison with the QPO frequencies observed in SGR 1900+14.
}
\label{fig:L1-1900-M14R12}
\end{figure}

\begin{figure}
\begin{center}
\includegraphics[scale=0.5]{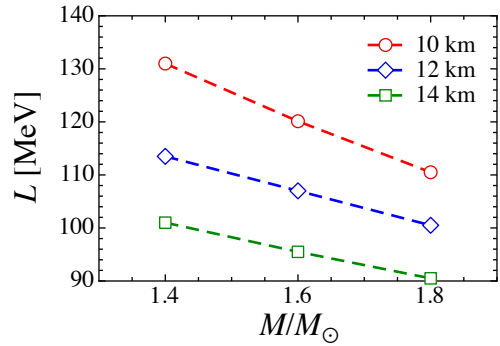} 
\end{center}
\caption{
(Color online) Same as Fig.\ \ref{fig:1806-L1}, but for
the best agreement with the QPO frequencies observed in SGR 1900+14.
}
\label{fig:1900-L1}
\end{figure}

     On the other hand, the low-lying QPOs observed in SGR 1900+14 can be 
similarly explained in terms of the fundamental frequencies of the shear 
torsional oscillations with $\ell=4$, 8, and 13.  For the stellar model with 
$M=1.4M_\odot$ and $R=12$ km, we show, in Fig.\ 
\ref{fig:L1-1900-M14R12}, the calculated frequencies as a function of $L$ and 
compare them with the QPO frequencies observed in SGR 1900+14. 
The best value of $L$ to reproduce the observed frequencies in SGR 
1900+14 is $L=113.5$ MeV, for which the calculated frequencies and 
relative errors are shown in Table \ref{tab:1900-1}. We can 
likewise obtain the best value of $L$ to reproduce the observed
QPO frequencies in SGR 1900+14 for each stellar model, which is plotted 
in Fig.\ \ref{fig:1900-L1}. Assuming again that the mass and 
radius of the oscillating neutron star are in the range 
$1.4\le M/M_\odot \le 1.8$ and 10 km $\le R \le$ 14 km, one can 
constrain $L$ as $90.5$ MeV $\le L \le 131.0$ MeV from the observed QPO 
frequencies of SGR 1900+14.

\begin{figure}
\begin{center}
\includegraphics[scale=0.5]{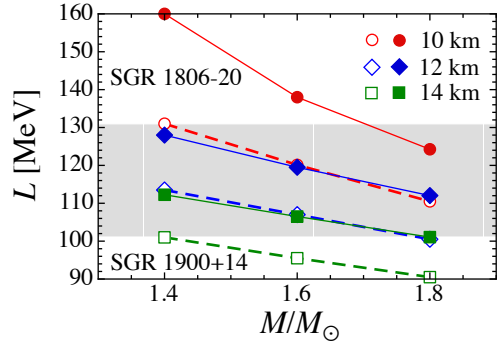} 
\end{center}
\caption{
(Color online) The values of $L$ (painted region) that are allowed 
simultaneously by both of the QPO observations in SGR 1806$-$20 and SGR 
1900+14 on the assumption that the corresponding neutron stars
have mass and radius in the range $1.4\le M/M_\odot\le 1.8$ and $10$ km 
$\le R\le$ 14 km.
}
\label{fig:L1}
\end{figure}

     Note that both observations in SGR 1806$-$20 and in SGR 1900+14 
should simultaneously be explained with a common value of $L$. We 
can thus obtain a more stringent constraint on $L$ as 101.1 MeV 
$\le L \le131.0$ MeV, which is shown in Fig.\ \ref{fig:L1}.  Some 
additional remarks are in order.  First, such constraint on $L$ in turn
would constrain the masses and radii of the central objects in SGR 1806$-$20 
and in SGR 1900+14 in such a way that the stellar models out of the 
painted region are ruled out.  Such mass-radius constraints would
be of particular use in the absence of empirical information about magnetar
masses and radii.  Second, within the nuclear model used here, as shown in 
\cite{OI2003}, the symmetry energy at density $n_0$, $S_0$, can be 
approximately written as a function of $L$:
\begin{equation}
   S_0 = 28~{\rm MeV} + 0.075 L. \label{eq:S0}
\end{equation}
With this relationship, the constraint on $L$ would translate into
$35.6$ MeV $\le S_0\le 37.8$ MeV.  One might thus be able to make 
severer constraints on $L$ and $S_0$ with the help of future possible
measurements of magnetar masses and/or radii.

%%%%%%%%%%%%%%%%%%%%%%%%%%%%%%%%%%%
% Figure 8
%%%%%%%%%%%%%%%%%%%%%%%%%%%%%%%%%%%
%%%%%%%%%%%%%%%%%%%%%%%%%%%%%%%%%%%

%%%%%%%%%%%%%%%%%%%%%%%%%%%%%%%%%%%
% Table 4
%%%%%%%%%%%%%%%%%%%%%%%%%%%%%%%%%%%
\begin{table*}
\centering
 \begin{minipage}{78mm}
\caption{Same as Table \ref{tab:1806-1}, but for the QPO frequencies
observed in SGR 1900+14.
}
\begin{tabular}{cccccc}
\hline\hline
 &  QPO frequency (Hz) & $\ell$ & ${}_0t_\ell$ (Hz) & relative
error (\%) & \\
\hline
  & 28 & 4 & 27.26 & 2.63 &  \\
  & 54 & 8 & 53.76 & 4.50 &  \\
  & 84 & 13 & 86.18 & $-2.60$ &  \\
\hline\hline
\end{tabular}
\label{tab:1900-1}
\end{minipage}
\end{table*}
%%%%%%%%%%%%%%%%%%%%%%%%%%%%%%%%%%%

%%%%%%%%%%%%%%%%%%%%%%%%%%%%%%%%%%%
% Figure 9
%%%%%%%%%%%%%%%%%%%%%%%%%%%%%%%%%%%
%%%%%%%%%%%%%%%%%%%%%%%%%%%%%%%%%%%

%%%%%%%%%%%%%%%%%%%%%%%%%%%%%%%%%%%
% Figure 10
%%%%%%%%%%%%%%%%%%%%%%%%%%%%%%%%%%%
%%%%%%%%%%%%%%%%%%%%%%%%%%%%%%%%%%%

\begin{figure}
\begin{center}
\includegraphics[scale=0.5]{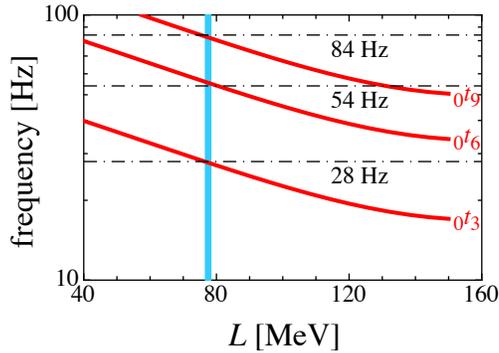} 
\end{center}
\caption{
(Color online) Alternative possible correspondence between the calculated 
fundamental frequencies of the shear torsional oscillations and the
QPO frequencies observed in SGR 1900+14. The vertical line denotes the 
value of $L$ that is consistent with the observations. 
}
\label{fig:L2-1900-M14R12}
\end{figure}

     Meanwhile, as an alternative possibility, one can identify the 
three QPOs observed in SGR 1900+14 as the $\ell=3$, 6, and 9 fundamental 
oscillations as in \cite{SW2009}.  As shown in Fig.\ \ref{fig:L2-1900-M14R12}, 
this identification is reasonable and indeed seems to be in better 
agreement with the observations than the former one as the $\ell=4$, 8, 
and 13 oscillations (see Table \ref{tab:1900-2}), although improvement
in the degree of agreement, which is small compared with the intrinsic errors 
of the fitting formula (\ref{eq:fit-0t2e}) used here, may not have to be 
taken seriously.  It is nonetheless interesting to note that the optimal
values of $L$ determined by the identification as the $\ell=3$, 6, and 9 
oscillations are in the range 58.0 MeV $\le L \le$ 89.5 MeV, as long as
one assumes the stellar models with $1.4 \le M/M_\odot \le 1.8$ and 10 km $\le
R\le$ 14 km.  This is because such values of $L$ are too small to become 
consistent with the optimal values of $L$ based on the identification of the 
four QPOs in SGR 1806$-$20 as the $\ell=3$, 4, 5, and 15 oscillations.

\begin{figure}
\begin{center}
\includegraphics[scale=0.5]{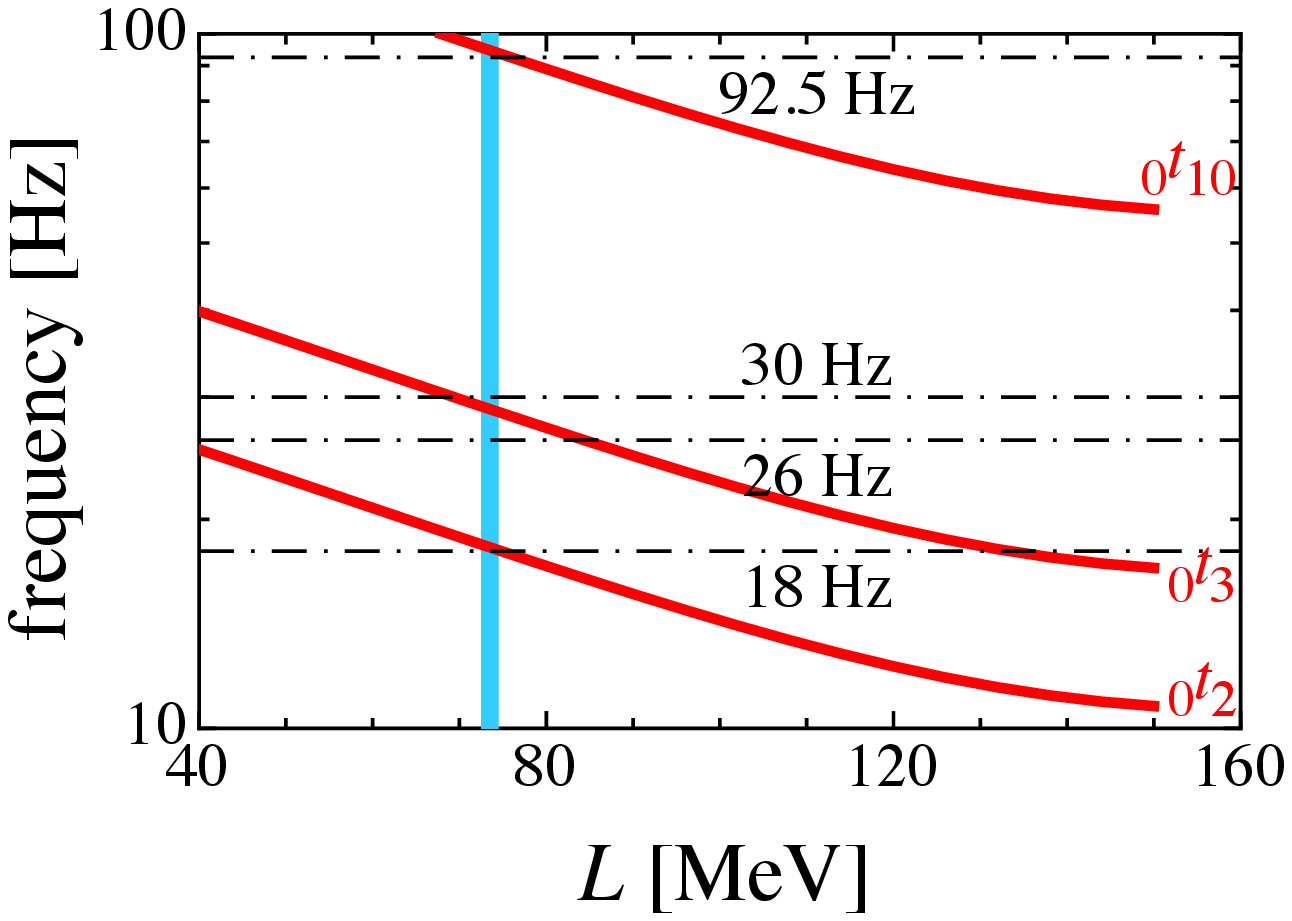} 
\end{center}
\caption{
(Color online) Alternative possible correspondence between the calculated 
fundamental frequencies of the shear torsional oscillations and the
QPO frequencies observed in SGR 1806$-$20 except 26 Hz.  The vertical line 
denotes the value of $L$ that is consistent with the observations.
}
\label{fig:L2-1806-M14R12}
\end{figure}

\begin{figure}
\begin{center}
\includegraphics[scale=0.5]{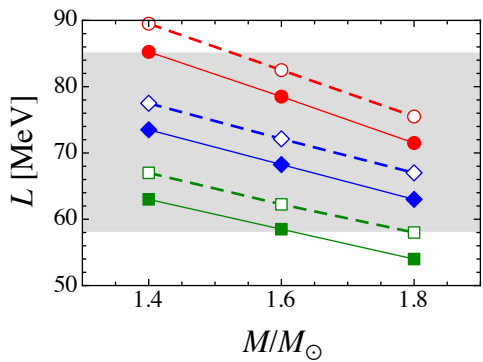} 
\end{center}
\caption{
(Color online) Same as Fig.\ \ref{fig:L1}, but for the alternative
identification.
}
\label{fig:L2}
\end{figure}

     To avoid such inconsistency, one could identify the 18, 30, and 
92.5 Hz QPOs in SGR 1806$-$20 as the $\ell=2$, 3, and 10 oscillations as 
in \cite{SW2009}, although the 26 Hz QPO in SGR 1806$-$20 remains 
to be identified.  In Fig.\ \ref{fig:L2-1806-M14R12}, we show how
such an identification works for a typical neutron star model with 
$M=1.4M_\odot$ and $R=12$ km; the corresponding relative errors are
listed in Table \ref{tab:1806-2}. By identifying the low-lying QPOs 
except the 26 Hz QPO in SGR 1806$-$20 as the $\ell=2$, 3, and 10 shear 
torsional oscillations, one can obtain the optimal values of $L$ as 54.0 MeV 
$\le L \le$ 85.3 MeV for the stellar models with $1.4 \le M/M_\odot \le$ 1.8 
and 10 km $\le R \le$ 14 km, which, as in Fig.\ \ref{fig:L2}, ensures the
presence of the allowed region of $L$ that simultaneously explain both 
observations in SGR 1806$-$20 and in SGR 1900+14, i.e., 58.0 MeV $\le L \le$ 
85.3 MeV.  This region corresponds to 32.4 MeV $\le S_0 \le$ 34.4 MeV via
Eq.\ (\ref{eq:S0}).  We emphasize that the present identification is 
again in better agreement with the observed frequencies except 
26 Hz than the former one as the $\ell=3$, 4, 5, and 15 oscillations.
This may invoke the possibility that the physics underlying the 26 
Hz QPO is missing.  We remark that various experimental constraints on $L$ 
seemingly favor smaller $L$, although they have yet to converge 
(see, e.g., Fig.\ 1 in \cite{NGWL2012} and also \cite{Tsang2012}).

%%%%%%%%%%%%%%%%%%%%%%%%%%%%%%%%%%%
% Figure 11
%%%%%%%%%%%%%%%%%%%%%%%%%%%%%%%%%%%
%%%%%%%%%%%%%%%%%%%%%%%%%%%%%%%%%%%

%%%%%%%%%%%%%%%%%%%%%%%%%%%%%%%%%%%
% Table 5
%%%%%%%%%%%%%%%%%%%%%%%%%%%%%%%%%%%
\begin{table*}
\centering
 \begin{minipage}{78mm}
\caption{Same as Table \ref{tab:1900-1}, but for the alternative
identification.
}
\begin{tabular}{cccccc}
\hline\hline
 &  QPO frequency (Hz) & $\ell$ & ${}_0t_\ell$ (Hz) & relative 
error (\%) & \\
\hline
  & 28 & 3 & 27.74 & 0.93 &  \\
  & 54 & 6 & 55.48 & $-2.74$ &  \\
  & 84 & 9 & 82.29 & 2.04 &  \\
\hline\hline
\end{tabular}
\label{tab:1900-2}
\end{minipage}
\end{table*}
%%%%%%%%%%%%%%%%%%%%%%%%%%%%%%%%%%%

%%%%%%%%%%%%%%%%%%%%%%%%%%%%%%%%%%%
% Figure 12
%%%%%%%%%%%%%%%%%%%%%%%%%%%%%%%%%%%
%%%%%%%%%%%%%%%%%%%%%%%%%%%%%%%%%%%

%%%%%%%%%%%%%%%%%%%%%%%%%%%%%%%%%%%
% Table 6
%%%%%%%%%%%%%%%%%%%%%%%%%%%%%%%%%%%
\begin{table*}
\centering
 \begin{minipage}{78mm}
\caption{Same as Table \ref{tab:1806-1}, but for the alternative
identification.
}
\begin{tabular}{cccccc}
\hline\hline
 &  QPO frequency (Hz) & $\ell$ & ${}_0t_\ell$ (Hz) & relative
error (\%) & \\
\hline
  & 18 & 2 & 18.23 & $-1.27$ &  \\
  & 26 & --- & --- & --- &  \\
  & 30 & 3 & 28.82 & 3.93 &  \\
  & 92.5 & 10 & 94.70 & $-2.38$ &  \\
\hline\hline
\end{tabular}
\label{tab:1806-2}
\end{minipage}
\end{table*}
%%%%%%%%%%%%%%%%%%%%%%%%%%%%%%%%%%%

%%%%%%%%%%%%%%%%%%%%%%%%%%%%%%%%%%%
% Figure 13
%%%%%%%%%%%%%%%%%%%%%%%%%%%%%%%%%%%
%%%%%%%%%%%%%%%%%%%%%%%%%%%%%%%%%%%

%%%%%%%%%%%%%%%%%%%%%%%%%%%%%%%%%%%%%%%%%%%%%%%%
\section{Conclusion}
\label{sec:V}
%%%%%%%%%%%%%%%%%%%%%%%%%%%%%%%%%%%%%%%%%%%%%%%%

   In this paper, we have investigated the fundamental frequencies of 
shear torsional oscillations in neutron star crusts, which contain
superfluid neutrons, for nine sets of the star's mass and radius 
as well as eleven models of the crust EOS, and considered possible 
relevance of the crustal shear modes to the QPOs observed from only a few 
SGRs in the afterglow of giant flares.  We have succeeded in identifying 
the low-lying QPOs as the shear modes of different $\ell$ in two ways (one
of which has an obvious caveat), leading to two separate allowed regions of 
the EOS parameter $L$ characterizing the density dependence of the symmetry 
energy via reasonable fitting of the calculated mode frequencies to the 
observed QPO frequencies.  The present results are basically the same as those 
obtained in our earlier publications \citep{SNIO2012,SNIO2013}, but the 
present work is more systematic and quantitatively finer.  We hope that future 
work in this direction, together with other empirical constraints on $L$, will 
eventually reduce the two allowed regions to one.

    In order to obtain a better constraint on $L$ asteroseismologically, 
however, many questions remain.  Among various properties of crustal matter 
that control the shear torsional oscillations, the shear modulus and the 
superfluid density play an especially significant role.  
%Estimates of the shear modulus in the Coulomb crystals of finite-size charges and in the pasta phases are thus desired, as well as of the superfluid density throughout a Wigner-Seitz cell.
Estimates of the shear modulus in the Coulomb crystals of
electron-screened, finite-size charges and in the liquid-crystalline
pasta phases are thus desired, as well as of the local superfluid density
throughout a Wigner-Seitz cell.
 For the bcc lattice of point charges, for example,
the electron screening acts to reduce the shear modulus \citep{HH2008,KP2013},
leading to decrease in the eigenfrequencies of the
shear torsional oscillations typically by 5 \%. Meanwhile, shell and pairing
effects on the nuclear charge, which are ignored in the present work,
are expected to have some consequence to the shear modulus
\citep{GMS2011,Deibel2013}.
Additionally, the crustal oscillations must be coupled with core magnetic
oscillations if one considers a magnetized neutron star. For instance, 
\cite{GCFMS2012a,GCFMS2012b} showed that the frequencies of the crustal
torsional oscillations can effectively increase, depending on the magnetic field
strength and structure (dipole-like poloidal, mixed toroidal-poloidal
with a dipole-like poloidal component and a toroidal field confined
to the region of field lines closing inside the star, and for
poloidal fields with an additional quadrupole-like component).
%They also showed that, with magnetic fields stronger than
%$\sim 10^{15}$ gauss, the magnetic oscillations become dominant and
%the crustal oscillations must disappear. If so, our constraint might make no sense.
On the other hand, it remains to be solved
how to deal with the junction conditions at the interface
between the crust and core regions, at least when one takes into account
the effect of superfluidity inside both core and crust regions.
%In that case, to reflect the effect of superfluidity on the junction condition,
%maybe, one should make an analysis with two-fluid model as in \cite{PL2013}.
Anyway we should also take into account the magnetic effect to obtain 
a better constraint on $L$, which will be addressed elsewhere.

%\newpage
%%%%%%%%%%%%%%%%%%%%%%%%%%%%%%%%%%%%%%%%%%%%%%%%
%\acknowledgments
%%%%%%%%%%%%%%%%%%%%%%%%%%%%%%%%%%%%%%%%%%%%%%%%
This work was supported in part by Grants-in-Aid for Scientific Research on 
Innovative Areas through No.\ 23105711, No.\ 24105001, and No.\ 24105008
and for the global COE program ``The Next Generation of Physics, Spun from 
Universality and Emergence" provided by MEXT, in part by Grants-in-Aid 
for Young Scientists (B) through No.\ 24740177 and for Research Activity 
Start-up through No.\ 23840038 provided by JSPS, and in part by the 
Yukawa International Program for Quark-hadron Sciences.

%\appendix
%%%%%%%%%%%%%%%%%%%%%%%%%%%%%%%%%%%%%%%%%%%%%%%%
%\section{}   % Appendix A
%\label{sec:appendix_1}
%%%%%%%%%%%%%%%%%%%%%%%%%%%%%%%%%%%%%%%%%%%%%%%%

%%%%%%%%%%%%%%%%%%%%%%%%%%%%%%%%%%%%%%%%%%%%%%%%


\begin{thebibliography}{999}
%%%%%%%%%%%%%%%%%%%%%%%%%%%%%%%%%%%%%%%%%%%%%%%%

\bibitem[\protect\citeauthoryear{Andersson \& Kokkotas}{1996}]{AK1996}
   Andersson N., Kokkotas K. D., 1996, Phys.\ Rev.\ Lett., 677, 4134

\bibitem[\protect\citeauthoryear{Andersson, Glampedakis \& Samuelsson}{2009}]{AGS2009}
   Andersson N., Glampedakis K., Samuelsson L., 2009, MNRAS, 396, 894

\bibitem[\protect\citeauthoryear{Chamel}{2012}]{Chamel2012}
   Chamel N., 2012, Phys.\ Rev.\ C, 85, 035801

\bibitem[\protect\citeauthoryear{Colaiuda \& Kokkotas}{2011}]{CK2011}
   Colaiuda A., Kokkotas K. D., 2011, MNRAS, 414, 3014

\bibitem[\protect\citeauthoryear{Colaiuda \& Kokkotas}{2012}]{CK2012}
   Colaiuda A., Kokkotas K. D., 2012, MNRAS, 423, 811

\bibitem[\protect\citeauthoryear{Deibel, Steiner, \& Brown}{2013}]{Deibel2013}  
   Deibel, A. T., Steiner, A. W., Brown, E. F., 2013, preprint (arXiv:1303.3270)

\bibitem[\protect\citeauthoryear{Demorest et al.}{2010}]{2Msun}  
   Demorest P. B., Pennucci T., Ransom S. M., Roberts M. S. E., Hessels J. W. T., 2010, Nature (London), 467, 1081

\bibitem[\protect\citeauthoryear{Gabler et al.}{2011}]{GCFMS2011}
   Gabler M., Cerd\'{a}-Dur\'{a}n P., Font J. A., M\"{u}ller E., Stergioulas N., 2011, MNRASL, 410, L37

\bibitem[\protect\citeauthoryear{Gabler et al.}{2012a}]{GCFMS2012a}
   Gabler M., Cerd\'{a}-Dur\'{a}n P., Stergioulas N., Font J. A., M\"{u}ller E., 2012a, MNRAS, 421, 2054
   
\bibitem[\protect\citeauthoryear{Gabler et al.}{2012b}]{GCFMS2012b}
   Gabler M., Cerd\'{a}-Dur\'{a}n P., Font J. A., Muller E., Stergioulas N., 2012b, preprint (arXiv:1208.6443)

\bibitem[\protect\citeauthoryear{Gearheart et al.}{2011}]{GNHL2011}
   Gearheart M., Newton W. G., Hooker J., Li B. A., 2011, MNRAS, 418, 2343

\bibitem[\protect\citeauthoryear{Grill, Margueron, \& Sandulescu}{2011}]{GMS2011}
   Grill F., Margueron J., Sandulescu  N., 2011, Phys.\ Rev.\ C, 84, 065801

\bibitem[\protect\citeauthoryear{Horowitz \& Hughto}{2008}]{HH2008}
   Horowitz C. J., Hughto J., preprint (arXiv:0812.2650)
   
\bibitem[\protect\citeauthoryear{Hurley et al.}{1999}]{H1999}
   Hurley K. et al., 1999, Nature, 397, L41

\bibitem[\protect\citeauthoryear{Iida \& Sato}{1997}]{IS1997}
   Iida K., Sato K., 1997, ApJ, 477, 294

\bibitem[\protect\citeauthoryear{Iida \& Baym}{2002}]{IB-II}
   Iida K., Baym G., 2002, Phys.\ Rev.\ D, 65, 014022

\bibitem[\protect\citeauthoryear{Kobyakov \& Pethick}{2013}]{KP2013}
   Kobyakov, D., Pethick, C. J., 2013, preprint (arXiv:1303.1315)

\bibitem[\protect\citeauthoryear{Kouveliotou et al.}{1998}]{K1998}
   Kouveliotou C. et al., 1998, Nature, 393, L235

\bibitem[\protect\citeauthoryear{Lattimer}{1981}]{L1981}
   Lattimer J. M., 1981, Annu.\ Rev.\ Nucl.\ Part.\ Sci., 31, 337

%\bibitem[\protect\citeauthoryear{Lattimer \& Lim}{2012}]{L2012}
%   Lattimer J. M., Lim Y., 2012, preprint (arXiv:1203.4286)

\bibitem[\protect\citeauthoryear{Lee}{2007}]{Lee2007}
   Lee U., 2007, MNRAS, 374, 1015

\bibitem[\protect\citeauthoryear{Levin}{2006}]{Levin2006}
   Levin Y., 2006, MNRASL, 368, L35

\bibitem[\protect\citeauthoryear{Levin}{2007}]{Levin2007}
   Levin Y., 2007, MNRAS, 377, 159

%\bibitem[\protect\citeauthoryear{Li, Chen \& Ko}{2008}]{LCK2008} 
%   Li B. A., Chen C. W., Ko C. M., 2008, Phys.\ Rep., 464, 113
   
\bibitem[\protect\citeauthoryear{Lorenz et al.}{1993}]{LRP1993}
   Lorenz C. P., Ravenhall D. G., Pethick C. J., 1993, Phys.\ Rev.\ Lett., 70, 379

%\bibitem[\protect\citeauthoryear{M\"{o}ller et al.}{2012}]{MMSY2012}
%   M\"{o}ller P., Myers W. D., Sagawa H., Yoshida S., 2012, Phys.\ Rev.\ Lett., 108, 052501

%\bibitem[\protect\citeauthoryear{Matsuo}{2013}]{Matsuo2013} 
%   Matsuo, M., 2013, private communication

\bibitem[\protect\citeauthoryear{Newton et al.}{2012}]{NGWL2012} 
   Newton W. G., Gearheart M., Wen D. H., Li B. A., 2012, preprint (arXiv:1212.4539)

\bibitem[\protect\citeauthoryear{Ogata \& Ichimaru}{1990}]{OI1990}
   Ogata S., Ichimaru S., 1990, Phys.\ Rev.\ A, 42, 4867

\bibitem[\protect\citeauthoryear{Okamoto et al.}{2012}]{Okamoto2012}
   Okamoto M., Maruyama T., Yabana K., Tatsumi T., 2012, Phys.\ Lett.\ B, 713, 284

\bibitem[\protect\citeauthoryear{Oyamatsu}{1993}]{O1993}
   Oyamatsu K., 1993, Nucl.\ Phys.\ A, 561, 431

\bibitem[\protect\citeauthoryear{Oyamatsu \& Iida}{2003}]{OI2003}
   Oyamatsu K., Iida K., 2003, Prog.\ Theor.\ Phys., 109, 631

\bibitem[\protect\citeauthoryear{Oyamatsu \& Iida}{2007}]{OI2007}
   Oyamatsu K., Iida K., 2007, Phys.\ Rev.\ C, 75, 015801

\bibitem[\protect\citeauthoryear{Passamonti \& Andersson}{2012}]{PA2012}
   Passamonti A., Andersson N., 2012, MNRAS, 419, 638

\bibitem[\protect\citeauthoryear{Passamonti \& Lander}{2013}]{PL2013}
   Passamonti A., Lander S. K., 2013, MNRAS, 429, 767

\bibitem[\protect\citeauthoryear{Pethick \& Potekhin}{1998}]{PP1998}
   Pethick C. J., Potekhin A. Y., 1998, Phys.\ Lett.\ B, 427, 7

\bibitem[\protect\citeauthoryear{Pethick, Chamel \& Reddy}{2010}]{PCR2010}
   Pethick C. J., Chamel N., Reddy S., 2010, Prog.\ Theor.\ Phys.\ Suppl., 186, 9

\bibitem[\protect\citeauthoryear{Ravenhall \& Pethick}{1994}]{RP1994}
   Ravenhall D. G., Pethick C. J., 1994, ApJ, 424, 846

\bibitem[\protect\citeauthoryear{Samuelsson \& Andersson}{2007}]{SA2007}
   Samuelsson L., Andersson N., 2007, MNRAS, 374, 256

\bibitem[\protect\citeauthoryear{Samuelsson \& Andersson}{2009}]{SA2009}
   Samuelsson L., Andersson N., 2009, Class.\ Quant.\ Gravity, 26, 155016

\bibitem[\protect\citeauthoryear{Sauls}{1989}]{Sauls}
   Sauls J. A., in {\it Timing Neutron Stars}, edited by \"{O}gelman H., van den Heuvel E.P.J. (Kluwer, Dortrecht, 1989), P. 457.

\bibitem[\protect\citeauthoryear{Schumaker \& Thorne}{1983}]{ST1983}
   Schumaker B. L., Thorne K. S., 1983, MNRAS, 203, 457

\bibitem[\protect\citeauthoryear{Sotani, Tominaga \& Maeda}{2001}]{Sotani2001}
   Sotani H., Tominaga K., Maeda K. I., 2001, Phys.\ Rev.\ D, 65, 024010

\bibitem[\protect\citeauthoryear{Sotani, Kohri \& Harada}{2004}]{Sotani2004}
   Sotani H., Kohri K., Harada T., 2004, Phys.\ Rev.\ D, 69, 084008

\bibitem[\protect\citeauthoryear{Sotani, Kokkotas \& Stergioulas}{2007}]{Sotani2007}
   Sotani H., Kokkotas K. D., Stergioulas N., 2007, MNRAS, 375, 261

\bibitem[\protect\citeauthoryear{Sotani, Kokkotas \& Stergioulas}{2008a}]{Sotani2008a}
   Sotani H., Kokkotas K. D., Stergioulas N., 2008a, MNRASL, 385, L5

\bibitem[\protect\citeauthoryear{Sotani, Colaiuda \& Kokkotas}{2008b}]{Sotani2008b}
   Sotani H., Colaiuda A., Kokkotas K. D., 2008b, MNRAS, 385, 2161

\bibitem[\protect\citeauthoryear{Sotani \& Kokkotas}{2009}]{Sotani2009}
   Sotani H., Kokkotas K. D., 2009, MNRAS, 395, 1163

\bibitem[\protect\citeauthoryear{Sotani}{2011}]{Sotani2011}
   Sotani H., 2011, MNRASL, 417, L70

\bibitem[\protect\citeauthoryear{Sotani et al.}{2011}]{SYMT2011}
   Sotani H., Yasutake N., Maruyama T., Tatsumi T., 2011, Phys.\ Rev.\ D, 83, 024014

\bibitem[\protect\citeauthoryear{Sotani et al.}{2012}]{SNIO2012}
   Sotani H., Nakazato K., Iida K., Oyamatsu K., 2012, Phys.\ Rev.\ Lett., 108, 201101

\bibitem[\protect\citeauthoryear{Sotani et al.}{2013a}]{SNIO2013}
   Sotani H., Nakazato K., Iida K., Oyamatsu K., 2013a, MNRASL, 428, L21

\bibitem[\protect\citeauthoryear{Sotani, Maruyama \& Tatsumi}{2013b}]{SMT2012}
   Sotani H., Maruyama T., Tatsumi T., 2013b, accepted for publication in Nucl. Phys. A

\bibitem[\protect\citeauthoryear{Steiner \& Watts}{2009}]{SW2009}
   Steiner A. W., Watts A. L., 2009, Phys.\ Rev.\ Lett., 103, 181101

\bibitem[\protect\citeauthoryear{Strohmayer et al.}{1991}]{SHOII1991}
   Strohmayer T., van Horn H. M., Ogata S., Iyetomi H., Ichimaru S., 1991, ApJ., 375, 679

\bibitem[\protect\citeauthoryear{Tsang et al.}{2012}]{Tsang2012} 
   Tsang M. B. {\it et al.}, 2012, Phys.\ Rev.\ C, 86, 015803

\bibitem[\protect\citeauthoryear{van Horn \& Epstein}{1990}]{vanhorn90}
   van Horn H. M., Epstein R. I., 1990, Bull.\ American Astron.\ Soc., 22, 748

\bibitem[\protect\citeauthoryear{Watts \& Strohmayer}{2006}]{WS2006}
   Watts A. L., Strohmayer T. E., 2006, Adv.\ Space Res., 40, 1446





\end{thebibliography}
\end{document}